\documentstyle[10pt,epsf,subfigure]{elsart}
  
\begin{document}
 
\pagenumbering{arabic} 
 
\newcommand{\ket}[1]{\left| #1 \right\rangle} 
\newcommand{\bra}[1]{\left\langle #1 \right|}  
\newcommand{\braket}[2]{\left\langle\left. #1 \right| \!#2\right\rangle} 
\newcommand{\matel}[3]{\left\langle  \left. \left. #1 \right| \!#2\right|#3 \right\rangle}  
%% derivee  
\newcommand{\dt}{{\partial _t}} 
\newcommand{\dx}{{\partial _x}} 
\newcommand{\dxx}{{\partial ^2_x}}  
\newcommand{\dtt}{{\partial ^2_t}} 
\newcommand{\dq}{{\partial _q}} 
\newcommand{\dpp}{{\partial _p}} 
%% begin et end 
\newcommand{\be}{\begin{equation}} 
\newcommand{\ee}{\end{equation}} 
\newcommand{\bc}{\begin{center}} 
\newcommand{\ec}{\end{center}} 
\newcommand{\benum}{\begin{enumerate}} 
\newcommand{\eenum}{\end{enumerate}} 
\newcommand{\bi}{\begin{itemize}} 
\newcommand{\ei}{\end{itemize}}  
\newcommand{\ba}{\begin{array}} 
\newcommand{\ea}{\end{array}} 
\newcommand{\bq}{\begin{quotation}} 
\newcommand{\eq}{\end{quotation}} 
%% mots revenant souvent 
\newcommand{\sch}{Schr\"odinger}
\newcommand{\wc}{Wick-Cutkowsky } 
\newcommand{\cst}{\mbox{cos}{\theta}} 
\newcommand{\sint}{\mbox{sin}{\theta}} 
\newcommand{\eps}{\sqrt{q^2+m^2}} 
\newcommand{\epsp}{\sqrt{q^{'2}+m^2}} 
\newcommand{\epsqx}{\sqrt{(q+x)^2+m^2}} 
\newcommand{\epsqxd}{\sqrt{1+ {2qx+x^2 \over q^2+m^2 }}} 
\newcommand{\cs}{\mbox{cos}} 
\newcommand{\si}{\mbox{sin}} 

\begin{frontmatter}
\title{Solutions of the Wick-Cutkosky model in the Light Front Dynamics}
\author{Mariane Mangin-Brinet, Jaume Carbonell}

\address{Institut des Sciences Nucl\'{e}aires
        53, Av. des Martyrs, 38026 Grenoble, France}

\begin{abstract}
We study relativistic effects
in a system of two scalar particles interacting via a scalar exchange in the Light Front 
Dynamcis framework.
The results are compared to those provided by Bethe-Salpeter and non relativistic equations.
It is found in particular that for massive exchange, the relativistic description
is of crucial importance even in the limit of zero binding energy.

\begin{flushleft}
\end{flushleft}
\end{abstract}

\end{frontmatter}

%%%%%%%%%%%%%%%%%%%%%%%%%%%%%%%%%%%%%%%%%%%%%%%%%%%%%%%%%%%%%%%%%%%%%%%%%%%%%%%
\section{Introduction}

Light Front Dynamics (LFD) is a field theoretically inspired hamiltonian approach specially 
well adapted for describing relativistic composite systems.
First suggested by Dirac \cite{D_49} it has been
since widely developed (see \cite{KP_91,COESTER_92,BUR_96,CDMK_98,Brodsky_98} 
and references therein)
and recently applied with success in its explicitly covariant version \cite{VAK_76} 
to the high momentum processes measured in
TJNAF \cite{CDMK_98,CK_99}.
In this approach the state vector is defined on a space-time hyperplane
given by $\omega\cdot x=\sigma$ where $\omega=(1,\hat{n})$ is a light-like four-vector.

We present here the first results obtained within this approach for the Wick-Cutkosky
model \cite{WC_54}.
This model describes the dynamics of two identical scalar particles of mass $m$ interacting
by the exchange of a massless scalar particle.
This first step towards more realistic systems constitutes an instructive case 
and is presently 
considered by several authors \cite{Tjon,Darewich,FSCS_98,Bakker,Lukas}. 
The model has been extended to the case where the
exchanged particle has non zero mass $\mu$ and used to build a 
relativistic scalar model for deuteron.

The results presented in this paper concern the S-wave bound states 
in the ladder approximation. 
They are aimed to
{\it (i)} compare the LFD and Bethe-Salpeter descriptions and study their non relativistic limits,
{\it(ii)}  disentangle the origin of the different relativistic effects,
{\it(iii)} evaluate the contribution of higher Fock components
and {\it(iv)} apply this study to a scalar  model for deuteron.

%#########################################################
\section{Equation for Wick-Cutkosky model}

We have considered the following lagrangian density:
\[ {\cal L}= {1 \over 2} \left( \partial_{\nu} \; \phi \partial^{\nu} \phi -m^2 \phi^{2}\right) + 
{1 \over 2} \left(\partial_{\nu} \chi \partial^{\nu} \chi -\mu^2 \chi^{2} \right)
- g \phi^2 \chi\]
where $\phi$ and $\chi$ are real fields.
In the case $\mu=0$ it corresponds to the Wick-Cutkosky model.
The wave function $\Psi$,  describing a bound state of two particles with  momenta $k_1$ and $k_2$, 
satisfies in the Light-Front the dynamical equation \cite{VAK_76,CDMK_98}
\be \label{lfd}
[4(q^2 +m^2)-M^2] \Psi(\vec{q},\hat{n})=-{m^2 \over 2 \pi^3} \int {d^3q' \over
\varepsilon_{q'} }V(\vec{q},\vec{q}\,',\hat{n},M^2)\Psi(\vec{q}\,',\hat{n})
\ee
Variable $\vec{q}$ is the
momentum of one of the particles in the reference system where $\vec{k_1}+\vec{k_2}=0$,
and tends in the non relativitic limit to the usual center of mass momentum.   
M represents the total mass of the composite system, $B=2m-M$
denotes its binding energy and $\varepsilon_q=\sqrt{q^2+m^2}$.
In the case of S-waves the wavefunction is a scalar quantity depending
only on scalars $q$ and $\hat{n}\cdot\vec{q}$ \cite{VAK_76,CDMK_98}.

The interaction kernel $V$ calculated in the ladder approximation is given by
\be \label{NoyK}
V(\vec{q},\vec{q}\,',\hat{n},M^2)= -{4 \pi \alpha \over Q^2+\mu^2}
\ee
with
\begin{eqnarray*}
Q^2&=&(\vec{q}-\vec{q'})^2-(\hat{n}\cdot\vec{q})(\hat{n}\cdot\vec{q'}) 
{ (\varepsilon_{q'}-\varepsilon_{q})^2 \over \varepsilon_{q'}\varepsilon_{q}}
+\left(\varepsilon_{q}^2+\varepsilon_{q'}^2-{M^2 \over 2}\right) 
\left|{\hat{n}\cdot\vec{q'} \over \varepsilon_{q'}}-{\hat{n}\cdot\vec{q} \over \varepsilon_{q}}\right|
\end{eqnarray*}
The coupling parameter $\alpha$ is defined by $\alpha=g^2 / 16\pi m^2$.
In the limit  $\varepsilon_b<<2m$ some
analytical solutions are known \cite{VAK_80}
and once removed the $\hat{n}$ dependence in the kernel 
-- formally setting $\hat{n}=0$ -- the LFD equation turns back to the Schr\"odinger
equation in the momentum space for the 
Yukawa or Coulomb potential, $\alpha$ being the usual fine structure constant.

\begin{figure}[htbp]
\begin{center}
\mbox{\epsfxsize=6cm \epsfysize=5cm \epsffile{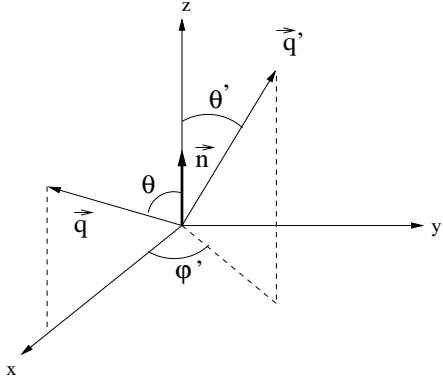}}
\end{center}
\caption{Kinematical variables. The $z$ axis is chosen along $\hat{n}$ and $\vec{q}$ can be 
restricted to the $xOz$ plane without loss of generality}\label{kinevar} 
\end{figure}

Equation (\ref{lfd}) has been solved with the coordinate choice displayed in
figure \ref{kinevar}. We have chosen $z$ axis along $\hat{n}$ and with no loss of generality $\varphi=0$.
The $\varphi'$ dependence of the 
kernel (\ref{NoyK}) can be performed analytically and (\ref{lfd}) turns into
the two dimensional integral equation
\be\label{lfd2}
[4(q^2 +m^2)-M^2]\Psi(q,\theta)={4m^2\alpha\over\pi}\int{q'^2\over\varepsilon_{q'}}dq'\sin\theta'd\theta'{1\over\sqrt{a^2-b^2}}\Psi(q',\theta') 
\ee
with 
\begin{eqnarray*}
a&=&q^2+q'^2- qq'
\left(2\cos\theta\cos\theta'+{(\varepsilon_{q'}-\varepsilon_q)^2\over\varepsilon_{q}\varepsilon_{q'}}\right)\cr
&+&\left(q^2+q'^2+2m^2 -{ M^2 \over 2}\right)   
\left|{q'\cos \theta' \over \varepsilon_{q'} }- {q \cos \theta \over \varepsilon_{q}}\right|  + \mu^2\\
b&=& 2qq'\si \theta \si \theta'        
\end{eqnarray*}

The kernel of (\ref{lfd2}) has an integrable singularity for $(q,\theta)=(q',\theta')$.
The equation is solved by expanding the solution on a spline functions basis
$S_i$, associated with coordinates $q$ and $\theta$:
$\Psi(q,\theta)=\sum_{i,j}c_{ij}S_i(q)S_j(\theta)$.
The r.h.s. two-dimensional integral is evaluated  using  Gauss
quadrature method adapted to treat the singularity. 
The unknowns of the problem are the coefficients $c_{ij}$, which are solutions of
a generalized eigeinvalue problem $\lambda BC=A(M^2)C$ for $M^2$ values such that $\lambda(M^2)=1$.

%%%%%%%%%%%%%%%%%%%%%%%%%%%%%%%%%%%%%%%%%%%%%%%%%%%%%%%%%%%%%%%%%%%%%%%%%%%%%%%
\section{Results}

\begin{figure}[hbtp] 
\begin{center}
\epsfxsize=7cm
\epsfysize=7cm{\mbox{\epsffile{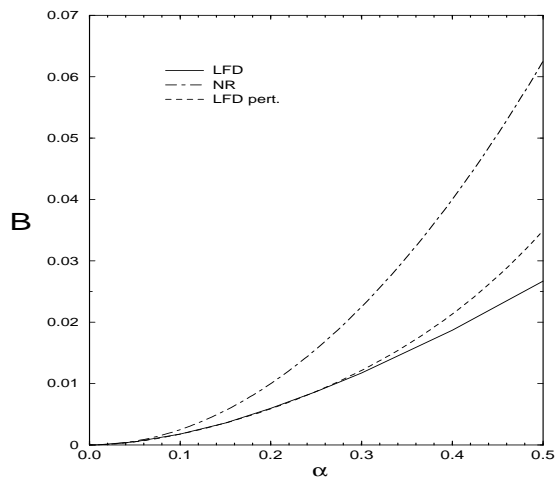}}} 
\caption{Binding energy (solid) for $\mu=0$ compared to non relativistic (dot-dashed) and 
perturbative (dashed) calculations}\label{Bmu0}
\end{center}
\end{figure}

The LFD binding energy for $\mu=0$ versus the coupling constant is displayed in figure  
\ref{Bmu0} (solid line). It is compared with the non relativistic values (dot-dashed line) and a 
first order perturbative calculation  $B_{pert}$ (dashed line),
valid also for Bethe-Salpeter (BS) equation \cite{FFT_73}, given by 
\be\label{bpert}
B_{pert}={m\alpha^2 \over 4}\left(1+{4 \over \pi }\alpha \log \alpha\right)
\ee 
Corresponding numerical values -- in $\hbar=c=m=1$ units -- are given in table \ref{tab_Balpha} together 
with the quantity $R={<q^2>\over m^2}$ usually used to evaluate the relativistic character of a system.
A first sight at this figure shows a significant departure from the non relativistic results already for $\alpha=0.1$.
This discrepancy -- which keeps increasing till $B$ reaches the maximum value of $2m$ -- is of 100\% for $\alpha=0.3$
whereas  $R$ remains very small. 
When evaluated using non relativistic solutions, $R$  is equal to $B$ (virial theorem),
what gives $R\approx 2\%$ for  $\alpha=0.3$, in contrast with the 100\%  effect in the binding.
The $R$ values obtained using the LFD solutions are even smaller (see table \ref{tab_Balpha}).
It is worth noticing the sizeable relativistic effects
observed in a system for which both the binding energy and the average momentum are small.

A good agreement with the perturbative calculation is found up to values $\alpha\approx 0.3$ where
the relative differences are $3\%$. 
The particular form of
equation (\ref{bpert}) ensures the existence of a non relativistic limit, the same for LFD and BS approaches, 
for the weakly bound states.
We will see later on, that this situation is particular to the case $\mu=0$.

\begin{table}
\caption{Binding energies for $\mu=0$ and $\mu=0.15$ as function of the coupling
constant $\alpha$}\label{tab_Balpha}
{\footnotesize
\begin{tabular}{l ccc ccc ccc cc}\hline  
$\alpha$             & 0.3  & 0.4   & 0.5   & 1.0   & 2.0   & 3.0  & 4.0   & 5.0  & 6.0  & 6.98 & 7.26 \\\hline
$10^2\,B_{\mu=0}$    & 1.17 & 1.87  & 2.67  & 7.68  & 21.0  & 38.0 & 58.5  & 84.0 & 118  & 200  &  -   \\
$10^2\,{<q^2>\over m^2}$            & 1.1  & 1.8   & 2.5   & 6.7   & 17    & 28   & 40    & 55   & 64   & 80   &  -   \\\hline 
$10^2\,B_{\mu=0.15}$ & -    & 0.190 & 0.570 & 4.36  & 16.5  & 32.7 & -     & 75.6 & 107  & 157  & 200 \\\hline
\end{tabular}
}
\end{table}

\begin{figure}[hbtp]
\begin{center}
\epsfxsize=6.7cm\epsfysize=6.5cm \subfigure[ ]{\epsffile{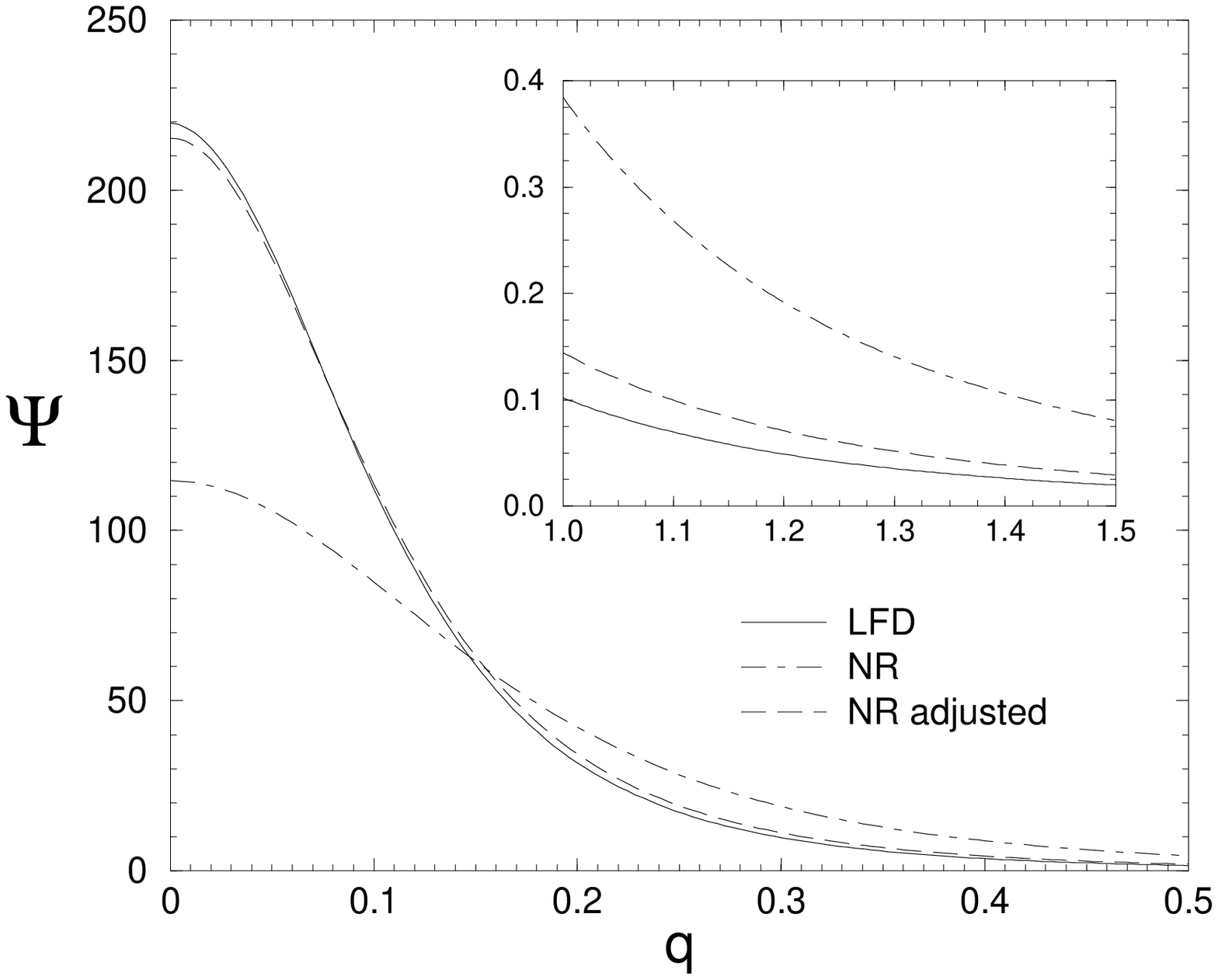}} \vspace{0.1cm}
\hspace{0.2cm}\epsfxsize=6.7cm\epsfysize=6.5cm \subfigure[ ]{\epsffile{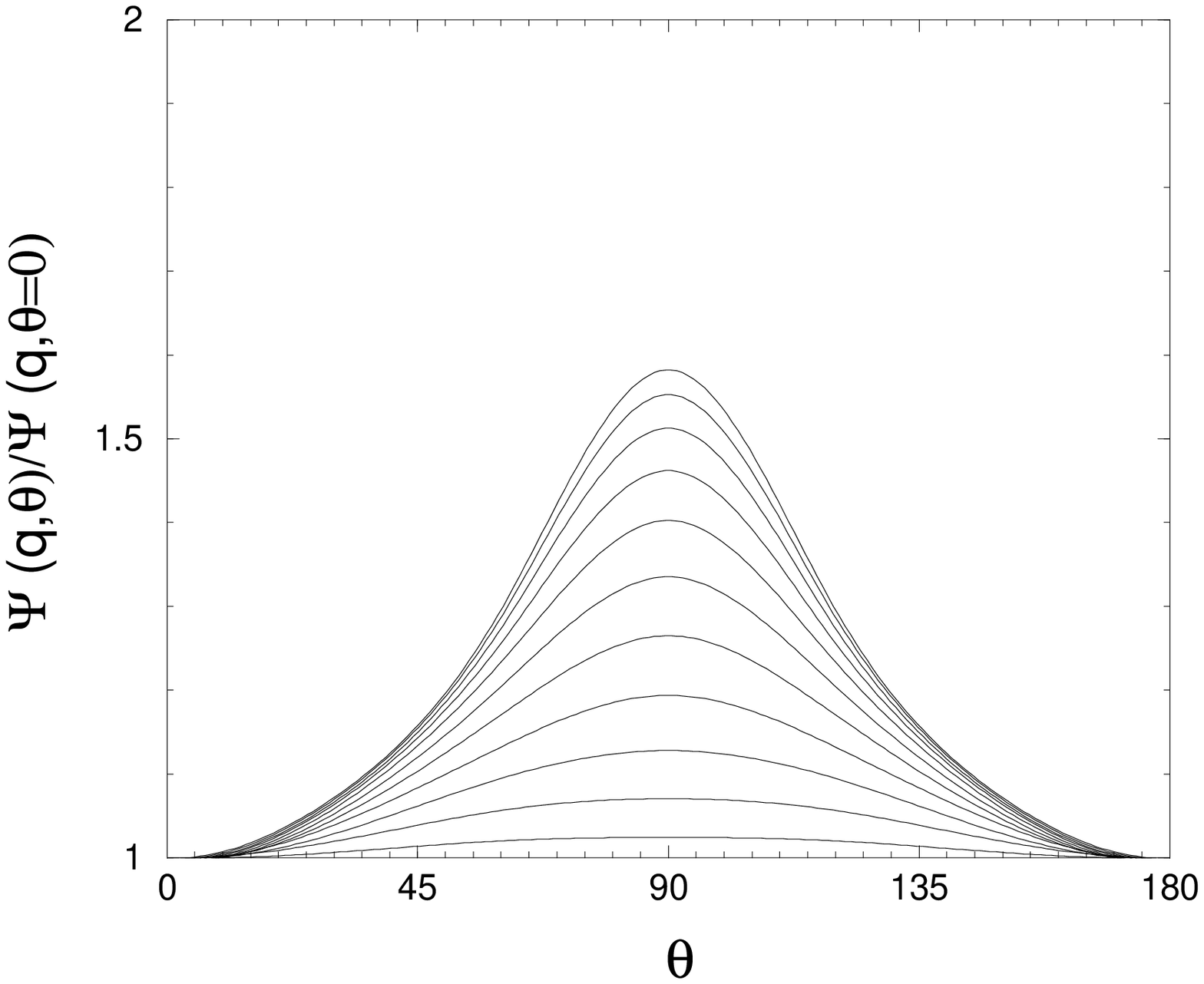}}
\vspace{-.5cm}
\caption{(a) S-wave LFD solution for $\theta=0$ (solid) compared with Coulomb wave function
with the same coupling constant (dot-dashed) and the same energy (dashed).
Figure (b) shows the angular dependence of the wave function for different values of $q$. 
Curves from
bottom to top correspond to increasing momenta from $q=0$ to $q=1.5$.}\label{wfS}
\end{center}
\end{figure}

The bound-state wavefunctions presented below are normalized according to
\be \label{norm}
{m \over (2 \pi)^3} \int \mid\Psi(q,\theta)\mid^2  \,{d^3q\over \varepsilon_q}  =1 
\ee
with $\varepsilon_q=m$ for the non relativistic case.
The LFD wave function $\Psi(q,\theta=0)$ obtained for $\alpha=0.5$
is compared in figure \ref{wfS}a (solid line)
with the corresponding non relativistic solution (dot-dashed line), that is Coulomb wave function.
The sizeable difference between both functions
is mainly due the differences in their binding energies: $B_{LFD}=0.0267$ 
whereas $B_{NR}=0.0625$ for the same coupling constant.
In order to compare wave functions with the same energy, the value of the coupling constant for the
non relativistic solution is adjusted to $\alpha_{NR}=0.327$.
The wave function obtained  (long-dashed curve) is then much closer to the 
relativistic one.

\vspace{0.0cm}

\begin{figure}[hbtp]
\begin{center}
\epsfxsize=8cm\epsfysize=8cm{\mbox{\epsffile{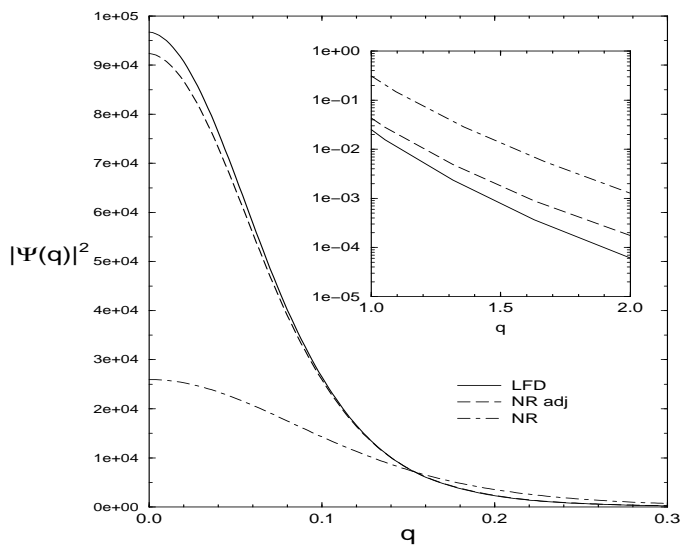}}} 
\caption{Squared modulus of the wave function integrated over $\theta$. 
The LFD solution (solid line) and the reajusted 
non relativistic one (dot-dashed line) have same binding energy.  
The small top right graph is a zoom of the high  momentum region.}\label{phik_int}
\end{center}
\end{figure}

Furthermore, in the region of high momentum transfer, 
the relativistic function is smaller than the Coulomb one,  
as expected from the natural cut-off of high momentum components introduced by relativity.
However, these differences can be accounted with the $\theta$ dependence 
of the LFD solutions, which exists even for S-waves. 
This angular dependence, normalized by the value of $\Psi(q,\theta=0)$, is shown in figure 
\ref{wfS}b
for different values of momentum $0\leq q\leq 1.5$. As one can see, the influence of the 
momentum orientation compared to the light-front plane is far from being negligible. 
This effect increases with $q$ and, for a fixed value of the momentum, 
is maximum when $\hat{q}\cdot \hat{n}=0$. 
For this kinematical configuration, i.e. relative momentum in the Light-Front plane,
the relativistic wave function $\Psi(q,\theta=90^o)$ at high momentum
is even found to be bigger than non relativistic one.
To get rid of this dependence, we compared $|\Psi(q,\theta)|^2$ integrated over the $\theta$
degree of freedom, both for relativistic and non relativistic solutions. 
The resulting functions, displayed in figure \ref{phik_int},
measure the effective relativistic effects in the wavefunctions.
At $q=0$ they remain at the level of $5\%$, once the energy is readjusted.  
In the high momentum region the 
relativistic solution is -- as expected -- smaller than the non relativistic one, but their
differences reach a factor three at $q=2$, and this for a moderate value of the coupling constant $\alpha=0.5$.

\begin{figure}[htb]
\begin{minipage}[t]{75mm}
\mbox{\epsfxsize=7.0cm\epsfysize=6.5cm\epsffile{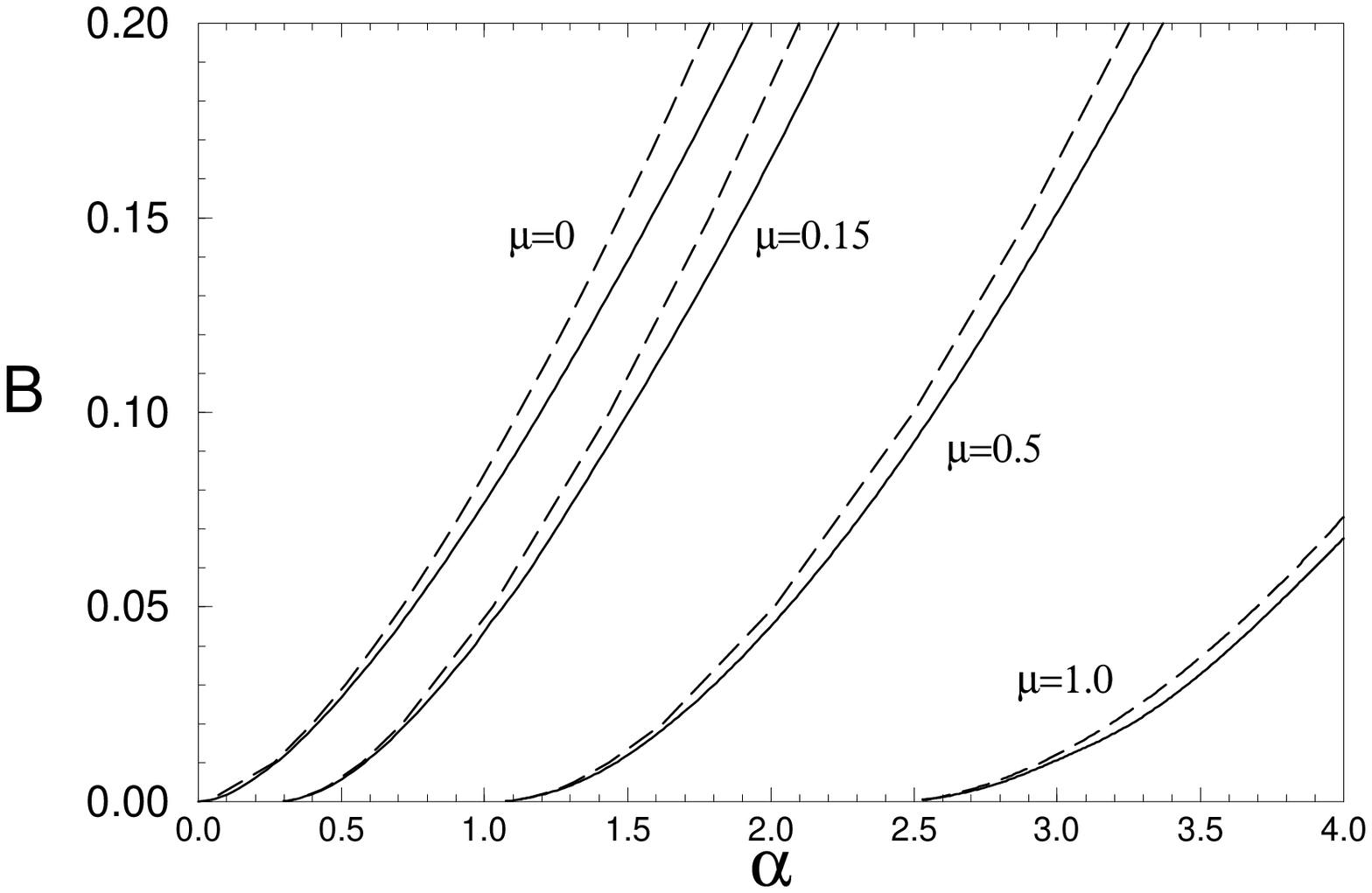}}
\caption{Binding energy as a function of $\alpha$ for different 
values of $\mu$ in LFD (solid) and BS (dashed) approaches}\label{B_LFD_BS_mu}
\end{minipage}
\hspace{0.1cm}
\begin{minipage}[t]{56mm}
\epsfxsize=5.4cm\epsfysize=6.0cm\mbox{\epsffile{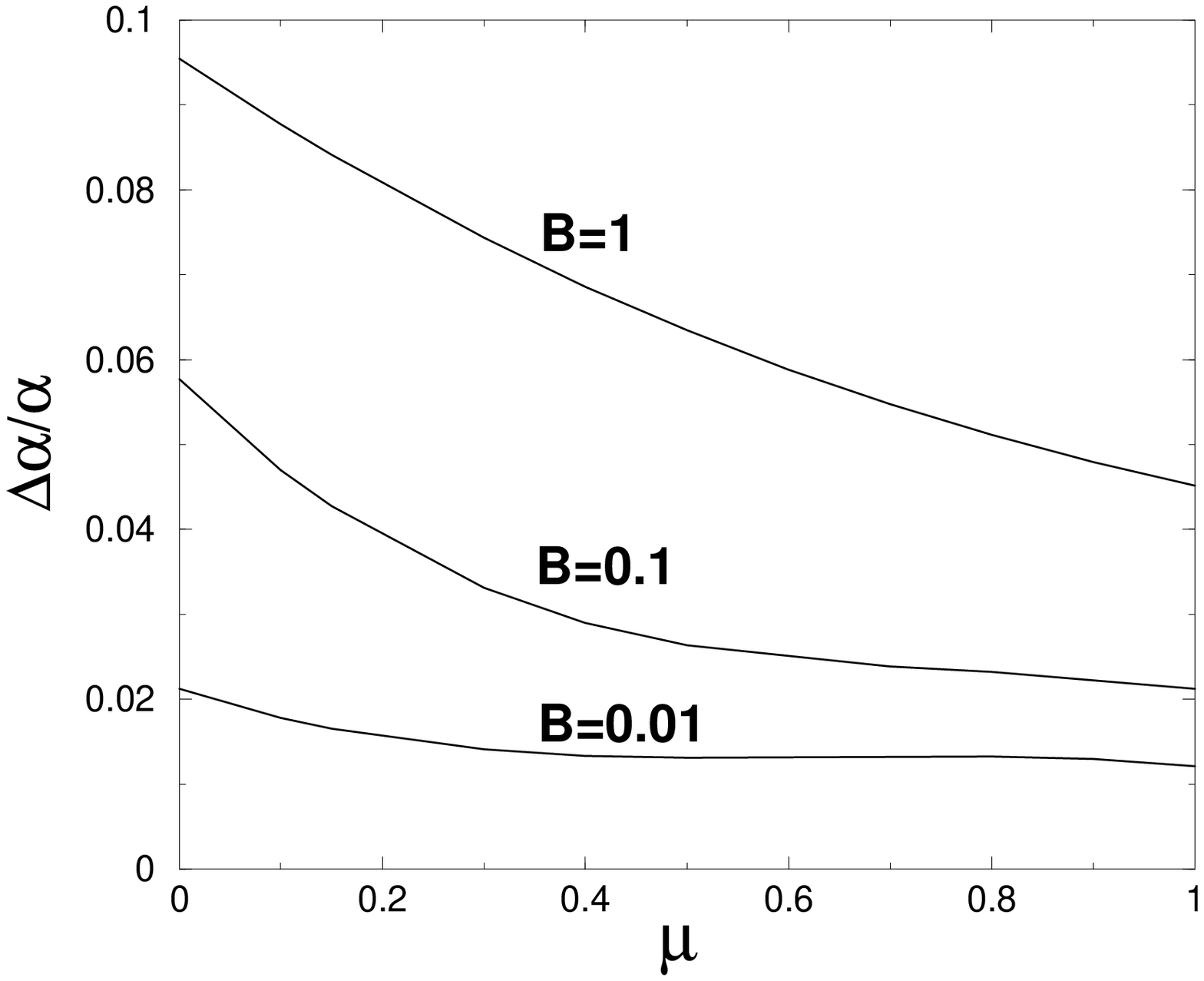}}
\caption{Differences in the coupling constant as a function of $\mu$ 
for  fixed values B.}\label{Dalpha_mu}
\end{minipage}
\end{figure}

In the case  $\mu\ne0$  there exists a critical $\alpha_0$ below which there is no bound solution.
Figure \ref{B_LFD_BS_mu} represents the binding energy $B$ as a function of the coupling
constant $\alpha$ for different values of $\mu$. 
They are compared with those provided by BS 
equation in the same ladder approximation,
whose kernel incorporates higher order intermediate states. 
We  have solved this equation using the method described in \cite{NT_FBS_96}. 
A first remark in comparing both approaches is that their results are seen to be close to each other.
This fact is far from being obvious -- 
specially for large values of coupling constant -- due to the differences in their ladder kernel.
A quantitative estimation of their spread can be given by looking into an horizontal
cut of figure \ref{B_LFD_BS_mu}, i.e. calculating the relative
difference in the coupling constant 
$(\alpha_{LFD}-\alpha_{BS})/\alpha_{LFD}$ for a fixed value 
of the binding energy.
The results, displayed in figure \ref{Dalpha_mu} for $B=1.0,0.1,0.01$,
show that relative differences 
({\it i})  are decreasing functions of $\mu$ for all values of $B$
({\it ii}) increase with $B$ but are limited to 10\% for the strong binding case $B=m$ 
which involves values of $\alpha\ge 5$.
This indicates the relatively weak importance of including 
higher Fock components in the ladder kernel even for strong couplings, as already discussed in \cite{Bakker}. 

\begin{figure}[hbtp]
\begin{center}
\epsfxsize=6.5cm\epsfysize=6.5cm \subfigure[ ]{\epsffile{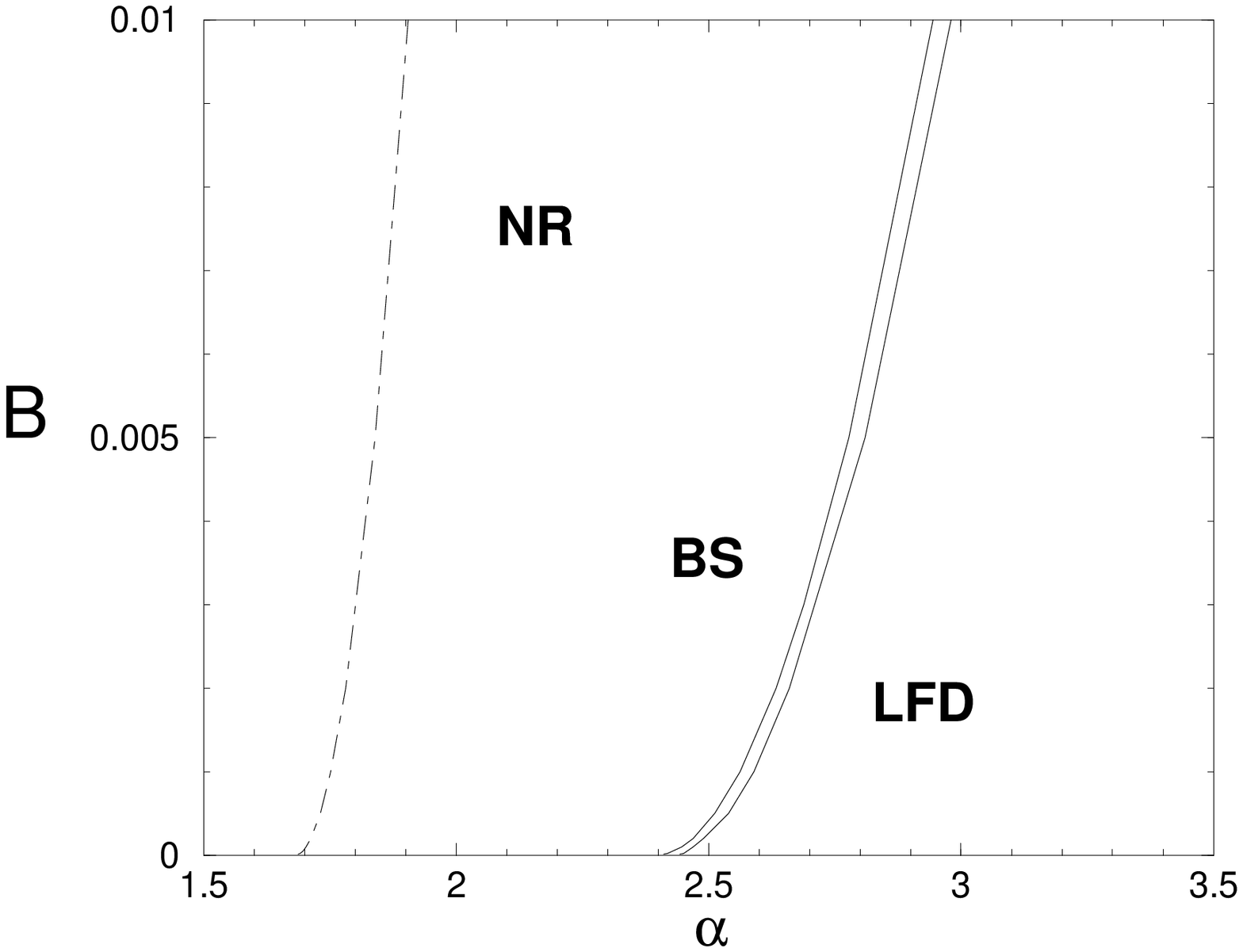}}\hspace{0.1cm}
\epsfxsize=6.5cm\epsfysize=7.3cm \subfigure[ ]{\epsffile{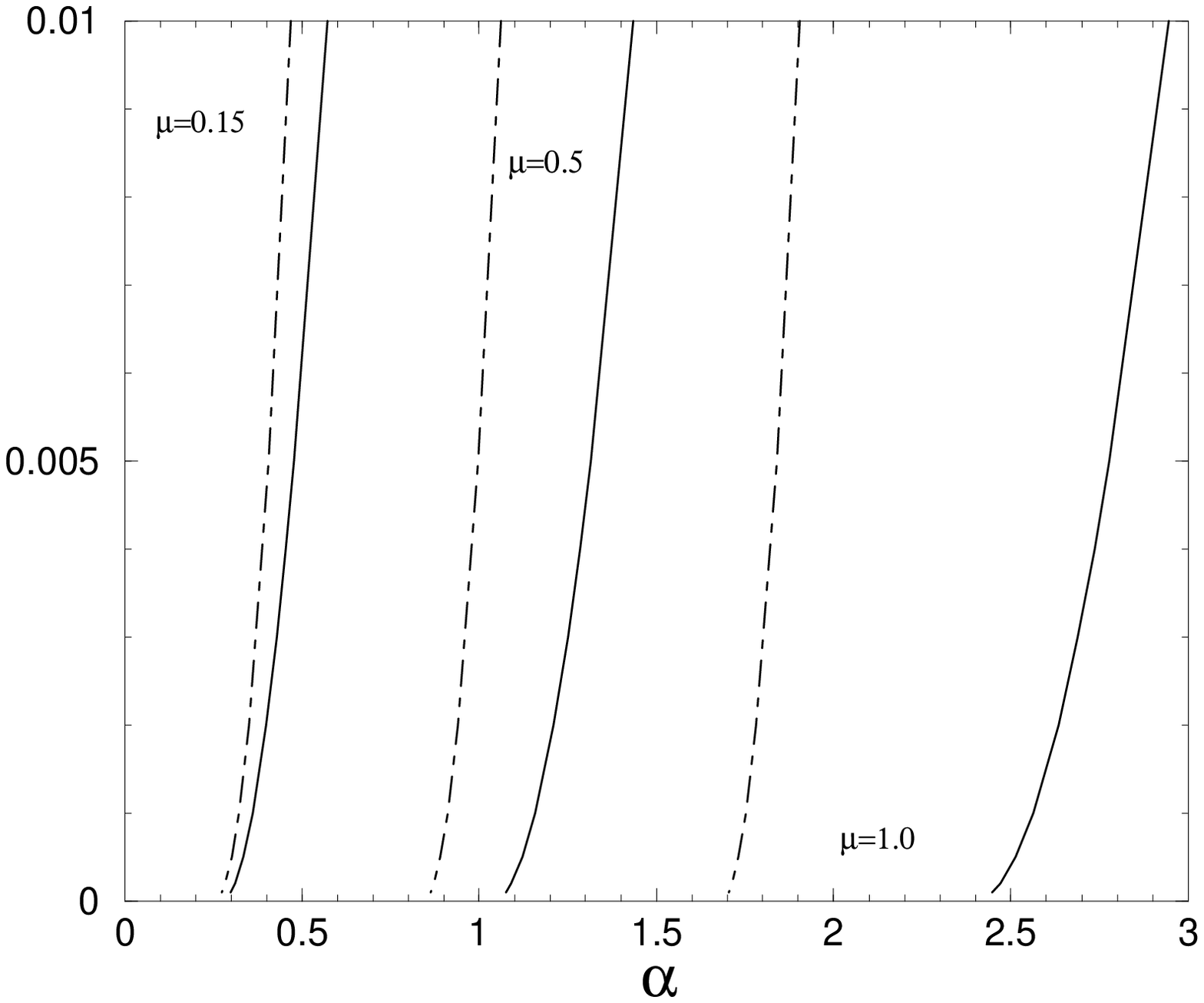}} 
\caption{Zero binding energy limit
of LFD and BS equations (solid) compared with the non relativistic  solutions (dot-dashed) 
for $\mu=1$ (a) and for different values of $\mu$ (b).}\label{B_alpha_LFD_BS_NR}
\end{center}
\end{figure}

It is interesting to
study the weak binding limit of both relativistic approaches and compare them
with the non relativistic calculations in the case $\mu\ne0$.
The results are given in figure \ref{B_alpha_LFD_BS_NR}a for $\mu=1$.
They show on one hand that LFD and BS (solid lines) 
converge towards very close, though slightly different, values
of the coupling constant (${\Delta\alpha\over\alpha}\approx 0.01$). 
On the other hand one can see, contrary to the $\mu=0$ case in figure \ref{Bmu0},
a dramatic departure of both relativistic approaches from a non relativistic theory (dot-dashed line),
even for negligibles values of binding energy.
The differences increase with
$\mu$ as shown in figure \ref{B_alpha_LFD_BS_NR}b in which LFD and BS results are not distinguished.
The origin of this departure
lies in the fact that the integral term in equation (\ref{lfd})
is dominated by the region $q'\sim\mu$, even for very small values of B,
and for the case $\mu\sim m$
the $q'\over m$ terms -- which make the difference between the non relativistic
and relativistic kernels -- are not longer negligible.
We conclude from that to the non adequacy of a non relativistic treatment
in describing systems interacting via massive fields,
what is the case of all the strong interaction physics when not described via gluon exchange.

\begin{figure}[ht] 
\begin{center}
\epsfxsize=8cm\epsfysize=8.cm\mbox{\epsffile{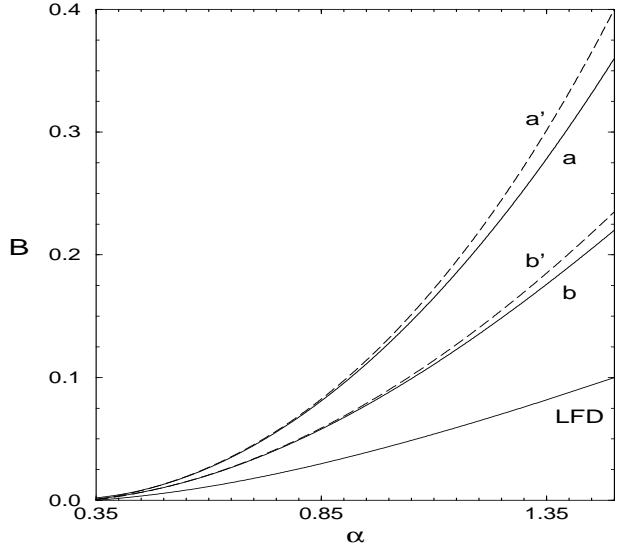}}
\caption{Relativistic contributions of different terms in equation (\ref{lfd}) (see text)} \label{Relcont}
\end{center}
\end{figure}

Some approximations of equation (\ref{lfd})
have been studied in order to disentangle the different contributions  
to the relativistic energies $B(\alpha)$ (see figure \ref{Relcont}). 
Equation (\ref{lfd}) is formally written $K\Psi={1\over\varepsilon_q}V\Psi$.
We first consider the case  of a non relativistic kernel $V$ -- i.e. a Yukawa potential --
and $\varepsilon_q=m$, with in curve $a$ the non relativistic kinematics $K=4 q^2+2mB$
and in curve $a'$  the relativistic one $K=4(q^2+m^2)-M^2$.
Curves $b$ and $b'$  are obtained in the same manner, but putting 
$\varepsilon_q=\sqrt{q^2+m^2}$. The last one  corresponds to the LFD equation. 
The results in figure \ref{Relcont} show that
the kinematical term $K$ has a very small influence on $B(\alpha)$, whereas
the contributions of $\varepsilon_q$ and $V$ to the total  binding are both essential.
We conclude from this study that the kinematical corrections alone, 
as they are performed e.g. in minimal relativity calculations,
are not representative of relativistic effects. Even by including them in the kernel
through $\varepsilon_q$ the results  obtained are wrong by a factor 2.

\begin{figure}[htbp] 
\begin{center}
\epsfxsize=8cm\epsfysize=7cm\mbox{\epsffile{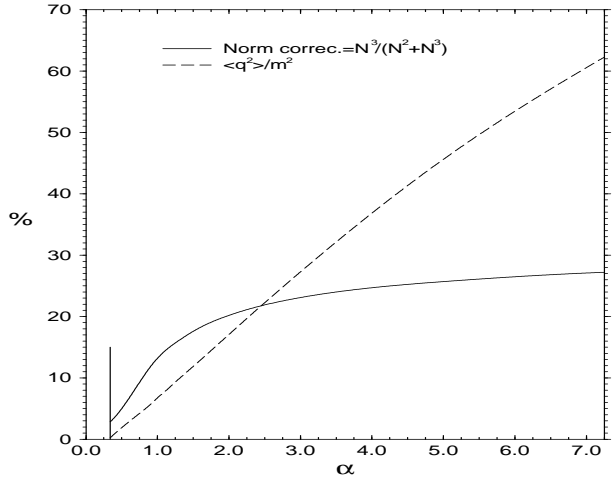}} 
\end{center}
\caption{Normalization correction due to 3-body Fock components and relativistic
parameter $R={<q^2>\over m^2}$ as functions of $\alpha$ for the case $\mu=0.15$.}\label{normal}
\end{figure}

In case of an energy dependent kernels
the normalization condition (\ref{norm}) is only approximate. 
This energy dependence denotes coupling to higher Fock components and the correct
normalization condition for the model considered reads $N^{(2)}+N^{(3)}=1$
where $N^{(3)}$ is the norm contribution from the three-body Fock component.
Using (\ref{norm}) only the two-body part is included.
One can show that the correction $N^{(3)}$ to the two-body normalization condition is given by:
\be
N^{(3)}=-{4m^2\over(2\pi)^6} \int
{d^3 q \over\varepsilon_q}{d^3 q'\over\varepsilon_{q'}} 
\Psi^{*}(q',\vec{q}\,'\cdot\hat{n})
{\partial V\over\partial M^2}(\vec{q},\vec{q}\,',\hat{n},M^2) \Psi(q,\vec{q}\cdot\hat{n})
\ee
This expression can be analytically integrated over two angles $\varphi$ and $\varphi'$ and we are let
with a four dimensional integration.
The three-body correction to the norm, i.e. the ratio ${N_{(3)}\over N_{(2)}+N_{(3)}}$,
as a function of the coupling constant is shown in figure \ref{normal} for the case $\mu=0.15$.
We remark that this correction is not zero at the critical value $\alpha=0.35$ 
corresponding to the $B=0$ threshold.
Its  behaviour in the region of large coupling tends asymptotically towards a value non exceeding $30\%$.
This is in contrast
with the evolution of the parameter $R$ introduced in the same figure to 
estimate the norm correction for a system with a given value of $R$. 
For deuteron, e.g., one has $R\approx 1\%$ and the expected normalization corrections are 
of the order of $4\%$.

%#########################################################
\section{A scalar model for Deuteron}

A simple relativistic model for deuteron in the LFD is obtained by adding
to the interaction kernel (\ref{NoyK}) a repulsive part exactly
analogous except for the sign of the coupling constant. Even if this 
procedure is no longer based on field theory -- a scalar exchange cannot produce a
repulsive interaction -- the potential obtained constitutes a LFD
relativistic version of the Malfliet and Tjon NN potential \cite{MT_69} on the form:
\be\label{MT_LFD}
V(\vec{q},\vec{q}\,',\hat{n},M^2)={V_R\over Q^2+\mu_R^2}-{V_A\over Q^2+\mu_A^2} 
\ee
The non relativistic model corresponds to $Q^2=(\vec{q}-\vec{q}\,')^2$ and
for the $^3S_1$ state the parameter set  
$V_R$=7.29, $V_A$=3.18, $\mu_A$=0.314 GeV, $\mu_R$=$2\mu_A$   
inserted in Schr\"odinger equation ensures a deuteron binding energy $B=2.23$ MeV. 

By solving the LFD equation (\ref{lfd}) with potential (\ref{MT_LFD}) we can estimate
the modification in the deuteron description due to a fully relativistic treatment.
The first result concerns its binding energy which becomes $B=0.96$ MeV.
The inclusion of relativity produces thus a dramatic repulsive effect, already drawn in \cite{DN_79}.
We emphasize that, as mentioned before in the case of Wick-Cutkosky model, 
the use of relativistic kinematics alone induces a very small change in the binding energy. 
The sizeable energy decrease  is almost entirely due to the r.h.s. part of (\ref{lfd}).  
To obtain a proper deuteron description in a relativistic frame 
it is necessary to readjust the parameters of the non relativistic model. 
A binding energy of 2.23 MeV is recovered 
with a repulsive coupling constant $\Lambda_R=6.60$ MeV
-- all other parameters being unchanged --
what represents an decrease of 10\% with respect to its original value. 
Another possibility is to increase 
the attractive coupling constant up to $V_A=3.37$. 
The relativistic effects in deuteron wave function 
depend sensibly on the way the energy is readjusted 
as well as on the relative angle $\theta$ between $\hat{n}$ and the momentum $\vec{q}$.
For instance when modyfing $V_A$ and for the value $\theta=0$,  
the zero of the relativistic wave function is shifted by $\approx$ 0.1 GeV/c
towards smaller values of $q$ and the differences
in the momentum region of $q=1.5$ GeV/c are 50\% in amplitude.

%%%%%%%%%%%%%%%%%%%%%%%%%%%%%%%%%%%%%%%%%%%%%%%%%%%%%%%%%%%%%%%%%%%%%%%%%%%%%%%
\section{Conclusion}

We have obtained the solutions for a scalar model in the Light Front Dynamics 
framework and in the ladder approximation. 
The results presented here concern the S-wave bound states.

We have found that the inclusion of relativity has a dramatic repulsive effect on binding energies
even for systems with very small ${<q>\over m}$ values. 
The effect is specially relevant when using a scalar model for deuteron: 
its binding energy is shifted from 2.23 MeV down to 0.96 MeV. This can be corrected
by decreasing of 10\%  the repulsive coupling constant, what indicates
the difficulty in determining beyond this accuracy the value of strong coupling constants 
within a non relativistic framework.  

Light-Front wave functions strongly differ from their non relativistic counterparts
if they are calculated using the same values of the coupling constant.
Once the interaction parameters are readjusted to get the same binding energy 
both solutions become closer but their differences are still sizeable.

The relativistic effets are shown to be induced mainly by the relativistic 
terms of the kernel. The
relativistic kinematics has only a small influence on the binding energy;
furthermore, its effect is attractive whereas 
the total relativistic one is strongly repulsive. 

The normalization corrections due to the three-body Fock components 
increase rapidly for small values of the coupling constant and saturate
at $\approx25\%$ in the ultra relativistic region. 
They have been estimated to $4\%$  in the deuteron. 

The LFD results are very close to those provided by Bethe-Salpeter equation for a wide range
of coupling constant despite the different physical input in their ladder kernel.
However in the case of systems interacting via a massive exchanged field, they both strongly differ 
from the non relativistic solutions even in the zero binding limit. 
This leads to the conclusion that
such systems cannot be properly described by using a non relativistic dynamics.

The case of higher angular momentum states for scalar particles requires
more formal developments and is presented in a forthcoming publication \cite{MCK_00}.

%%%%%%%%%%%%%%%%%%%%%%%%%%%%%%%%%%%%%%%%%%%%%%%%%%%%%%%%%%%%%%%%%%%%%%%%%%%%%%%%%%%%%%%%%%%%%%%%%%%%%%
\ack{We are grateful to V.A. Karmanov for enlightening discussions during this work
and for a careful reading of the manuscript. We thank L. Theussl for providing 
some results from \cite{Lukas} prior to its publication}

\end{document}